\newif\ifsubmode
\submodefalse

\ifsubmode
  \documentclass[12pt,preprint]{aastex}  %% For preprint style with 12pt type
  \received{}
  \revised{}
  \accepted{}
\else
  \documentclass{emulateapj}  %% To emulate ApJ style
  \slugcomment{Accepted for publication in ApJ Letters}
\fi

\shortauthors{Fassnacht et al.}
\shorttitle{GOODS Strong Gravtiational Lens Search}

\begin{document}

\title{Strong Gravitational Lens Candidates in the GOODS ACS Fields\altaffilmark{1}}

\author{C. D. Fassnacht}
\affil{
   Department of Physics,
   University of California,
   1 Shields Avenue,
   Davis, CA 95616
}
\email{fassnacht@physics.ucdavis.edu}

\and

\author{L. A. Moustakas, S. Casertano, H. C. Ferguson, R. A. Lucas,
 Y. Park\altaffilmark{2}}
\affil{
   Space Telescope Science Institute,
   3700 San Martin Drive, 
   Baltimore, MD 21218}
\email{leonidas, stefano, ferguson, lucas, ypark@stsci.edu}

\altaffiltext{1}{Based on observations taken with the NASA/ESA 
{\em Hubble Space 
Telescope}, which is operated by the Association of Universities
for Research in Astronomy, Inc.\ (AURA) under NASA contract
NAS5--26555}
\altaffiltext{2}{Also 
   Department of Physics and Astronomy,
   Johns Hopkins University,
   Baltimore, MD 21218
}

\begin{abstract}
We present results from a systematic search for strong gravitational
lenses in the GOODS ACS data.  The search technique involves creating
a sample of likely lensing galaxies, which we define as massive
early-type galaxies in a redshift range $0.3<z<1.3$.  The target
galaxies are selected by color and magnitude, giving a sample of 1092
galaxies.  For each galaxy in the sample, we subtract a smooth
description of the galaxy light from the $z_{850}$-band data.  The
residuals are examined, along with true-color images created from the
$B_{435}V_{606}i_{775}$ data, for morphologies indicative of strong
lensing.  We present our six most promising lens candidates, as
well as our full list of candidates.  

\end{abstract}

%\keywords{
%   galaxies: general ---
%   galaxies: high-redshift ---
%   gravitational lensing ---
%   surveys
%}

\section{Introduction}

A great strength of gravitational lenses as astrophysical tools is
that they provide direct measures of total mass, without the
requirement that the mass be luminous or baryonic.  At the most basic
level, the image separation in lens systems provides a nearly
model-independent measurement of the projected mass of the lensing
galaxy \citep[e.g.,][]{csklensmass,wplensmass}.  
The combination of additional observational constraints and more
sophisticated modeling can provide data on the elongation and
orientation of the mass distribution \citep[e.g.,][]{keetonshear}, the
radial mass profile of the lensing galaxy
\citep[e.g.,][]{kt0047,rusin1152,cskrings,wucknitz0218}, and the
possible existence of dark matter subhaloes
\citep[e.g.,][]{substruct,cdmstruct1,cdmstruct3,cdmstruct5,cdmstruct4}.
Therefore, gravitational lenses provide a wealth of information about
moderate redshift galaxies and can provide important insights into
galaxy evolution \citep[e.g.,][]{keetongalevol,rusingalevol}.
An increase in the number of known lenses benefits almost all these
studies and thus motivates searches for new lenses.  The data obtained
with the Advanced Camera for Surveys (ACS) 
on the {\em Hubble Space Telescope} as part of the Great Observatories
Origins Deep Survey \citep[GOODS;][] {goods} offers an unprecedented
combination of angular resolution, depth, and sky coverage.  In this
paper we report the results of a systematic search for lenses in the
data obtained in the first three epochs of observations of the GOODS fields.

\section{The target population}

The search strategy that we used was focused on the likely lens
galaxies rather than the likely background sources.  This is in
contrast to the searches conducted at radio wavelengths
\citep[e.g.,][]{class1,class2,winnsearch} and optical searches
targeting known quasars \citep[e.g.,][]{hstlens}.  For this search, we
selected the objects most likely to produce strong lenses, namely
massive early-type galaxies at $0.3<z<1.3$.  Theoretical studies
indicate that these galaxies should dominate the lensing population,
with only $\sim$10\% or fewer of strong lenses produced by spiral
galaxies \citep[e.g.,][]{tog,ftlensrate,maozrix}.  Similarly, a
ray-tracing analysis conducted on the northern Hubble Deep Field data
\citep[HDF-N;][]{hdfn} indicated that nearly all the lensing
cross-section was provided by massive ellipticals at $z \sim 1$
(Blandford, Surpi, \& Kundi\'c 2001).  Thus, our target sample was chosen to
maximize the chance of finding massive early-type galaxies in the
desired redshift range.

All candidate selection was done using the ACS data obtained in the
F435W, F606W, F775W, and F850LP filters (hereafter, $B_{435}$,
$V_{606}$, $i_{775}$, and $z_{850}$).  We consider only objects with
isophotal magnitudes $z_{850,iso}<23.0$.  Colors were measured in
matched apertures, with the apertures defined by the $z_{850}$-band
data\footnote{For complete details of the source extraction,
photometry, and limiting magnitudes, see \citet{goods}.}.  In a
color-color plot of $(B_{435}-i_{775})$ vs $(i_{775}-z_{850})$,
galaxies at $z > 0.3$ with old stellar populations should lie in the
upper part of the diagram (see Figure 1).  We define our target region as
$(B_{435}-i_{775}) > 2.0$, which produces a sample of 1092 galaxies.
Figure 1 also contains, as examples, no-evolution tracks for
different types of galaxies, which show solely the effect of spectral
features moving into different filter bands for galaxies at different
redshifts.  Although the track for elliptical galaxies in Figure 1
would suggest that our target region contains ellipticals in a
wider redshift range than desired, the combination of the low volume
probed at low redshifts, cosmological dimming at high redshifts, and
evolution implies that the redshift range selected is closer to the
desired $0.3<z<1.3$.  Based on comparisons with both non-evolving and
passive evolution models
\citep{kinneysed}, as well as spectroscopy of $\sim200$ galaxies in the
southern GOODS field, we find that the regions marked in the figure do
contain the vast majority of the massive early-type galaxies in the
relevant redshift interval.  The target region also includes a large
contamination ($\sim20$\%) from highly-obscured systems (particularly
dusty bulge-dominated galaxies).  Since our goal is to be as inclusive
as possible, without needing to examine the full catalog 
% ($\sim$3000 galaxies brighter than $z_{iso}=23$) 
in detail, this gives an adequate sample of target galaxies.

\ifsubmode
\else
  \begin{figure}
  \plotone{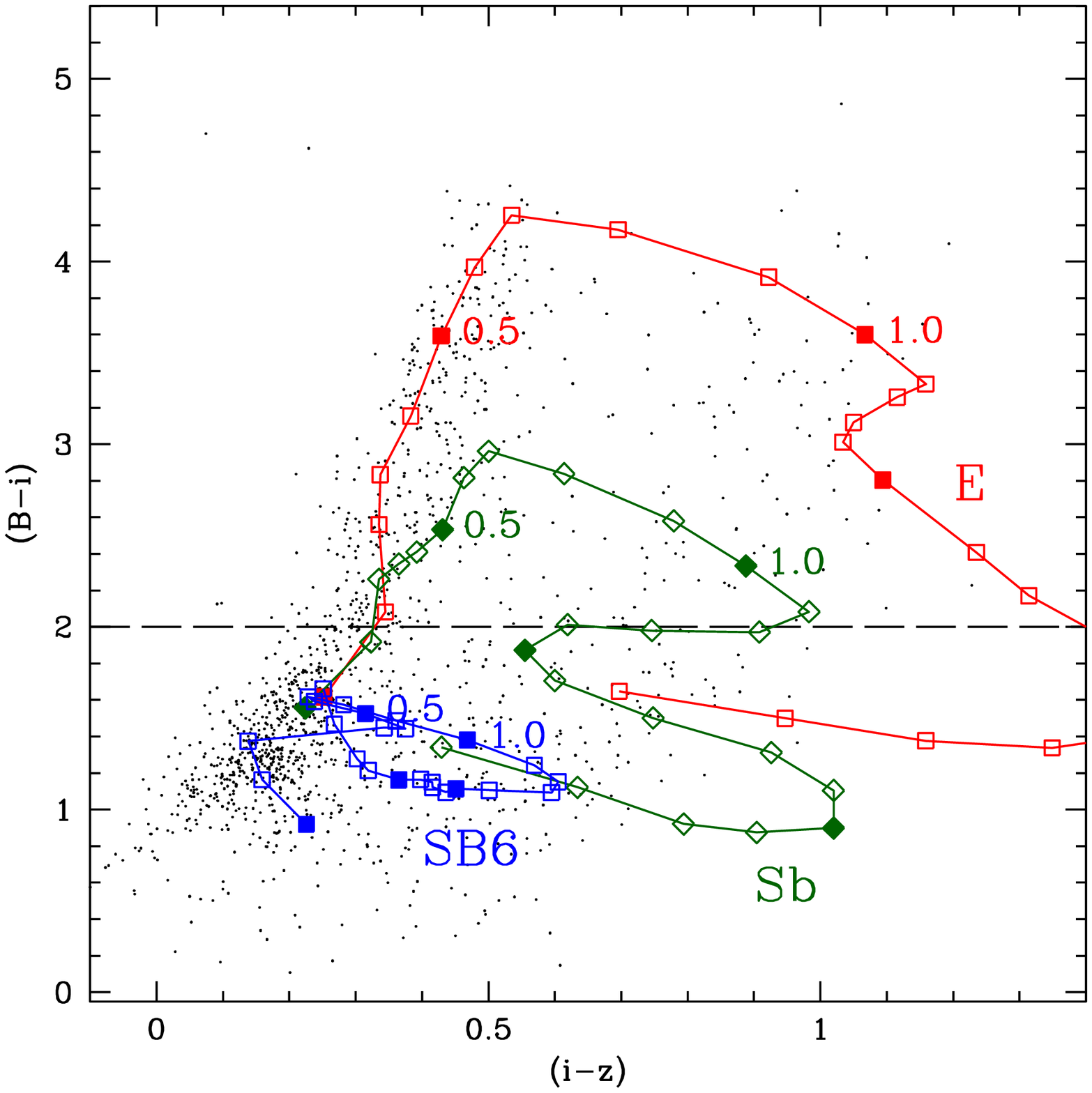}
  \caption{
   Color-color diagram for galaxies in the GOODS southern field
   $z_{850}$-selected catalog, limited to $z_{850,iso}<23.0$.  The target
   galaxies for the lens search lie above the dashed line.  The open
   symbols connected by lines represent the expected change with redshift
   of colors for non-evolving elliptical (E), Sb, and starburst (SB6)
   galaxies.  The points are separated by $\Delta z = 0.1$, starting at
   $z = 0$ and with solid points every $\Delta z = 0.5$.  On each track,
   the points corresponding to $z=0.5$ and $z=1.0$ are marked.
   % The plot for the northern GOODS field is qualitatively similar.
  \label{fig_ellselect}}
  \end{figure}
\fi

\section{Lens-search techniques}

\subsection{Subtraction of galaxy models}

Our primary technique for searching the target sample for
gravitational lens candidates was to model the light distribution of
the target galaxy and then subtract the model from the data.  The
residuals for each target were examined for indications of
gravitational lensing.  To increase the robustness of our results, we
used two methods to subtract the galaxy emission for each target.  

The first method used customized IDL scripts that provided an
empirical description of the target galaxy emission.  The galaxy
centroid, ellipticity, and position angle were calculated from the
light distribution.  The input image was split into a series of
concentric ellipses, each with the input ellipticity and position
angle.  The empirical description of the galaxy emission was then
produced by boxcar-smoothing the distribution of flux as a function of
semi-major axis.  This simple approach provided acceptable
representations of most of the target galaxies, and non-symmetric
features were clearly visible in the residual images produced by
subtracting the smoothed image from the input.  To judge the
effectiveness of the technique in subtracting smoothly-distributed
galaxy light, we use as a figure of merit the ratio of the counts in
the residual image to the counts in the original image, summed within
a 3\arcsec-diameter aperture.  For $\sim$75\% of the targets, this
ratio was less than 0.05, indicating a good fit to the galaxy emission.
All targets producing higher ratios had
either close neighbors or the kind of asymmetric structure that the
routine was designed to find, e.g., knots of star formation, bars,
dust lanes, possible lensed images, etc.
Thus, this method is sensitive to compact lensed images and to
tangentially stretched arcs.  However, complete or nearly complete
Einstein rings would be subtracted away by this method.  Therefore, we
used a second galaxy-subtraction method to complement the empirical
technique.

The second technique involved fitting parametric models to all
candidate lens galaxies.  Experimentation with various combinations of
parametric laws \citep[using GALFIT;][]{galfit} showed that a single
Sersic fit did remarkably well for the majority of galaxies in our
sample, with a median $\chi^2_{fit}/\nu\approx0.9$.  In most of the
cases, the residuals produced by the GALFIT subtraction did not differ
qualitatively from those produced by the empirical method, and no
believable lens candidates were found solely from an examination of
the GALFIT residuals.

\subsection{Selection Criteria}

The residuals produced by the galaxy-subtraction methods were
carefully examined, as were the unsubtracted true-color
($B_{435}V_{606}i_{775}$) images of the target galaxies.  The latter
provided invaluable color information about features in the target
images, which was incorporated into the candidate selection process.
The criteria used to select lens candidates were based on features
commonly seen in HST images of strong gravitational
lenses\footnote{The CfA-Arizona Space Telescope Lens Survey web site,
at \url{http://cfa-www.harvard.edu/castles/} provides an excellent
compilation of such images.}.  Any one of the following features seen
in either the residual images or in the true-color images would cause
a target to be flagged as a possible lens candidate: (1) two compact
images on opposite sides of the target galaxy; (2) multiple (more than
two) compact images arranged in a typical lensing morphology, (3) an
arc, possibly with a faint counterimage, tangentially stretched
relative to the target galaxy, especially if the arc is bluer than the
galaxy, (4) a ring of possibly blue emission surrounding the target
galaxy.  Because gravitational lensing is an achromatic phenomenon,
the alleged lensed images should have the same colors.  However, some
leeway was allowed to incorporate the possible effects of differential
extinction \citep[e.g.,][]{rdb1608,falcodust}.
Two of the authors independently examined all of the target galaxies.  The
list presented in Table~\ref{tab_allcand} is the union of the two 
selections.

\ifsubmode
\else
  \begin{figure*}
  \plotone{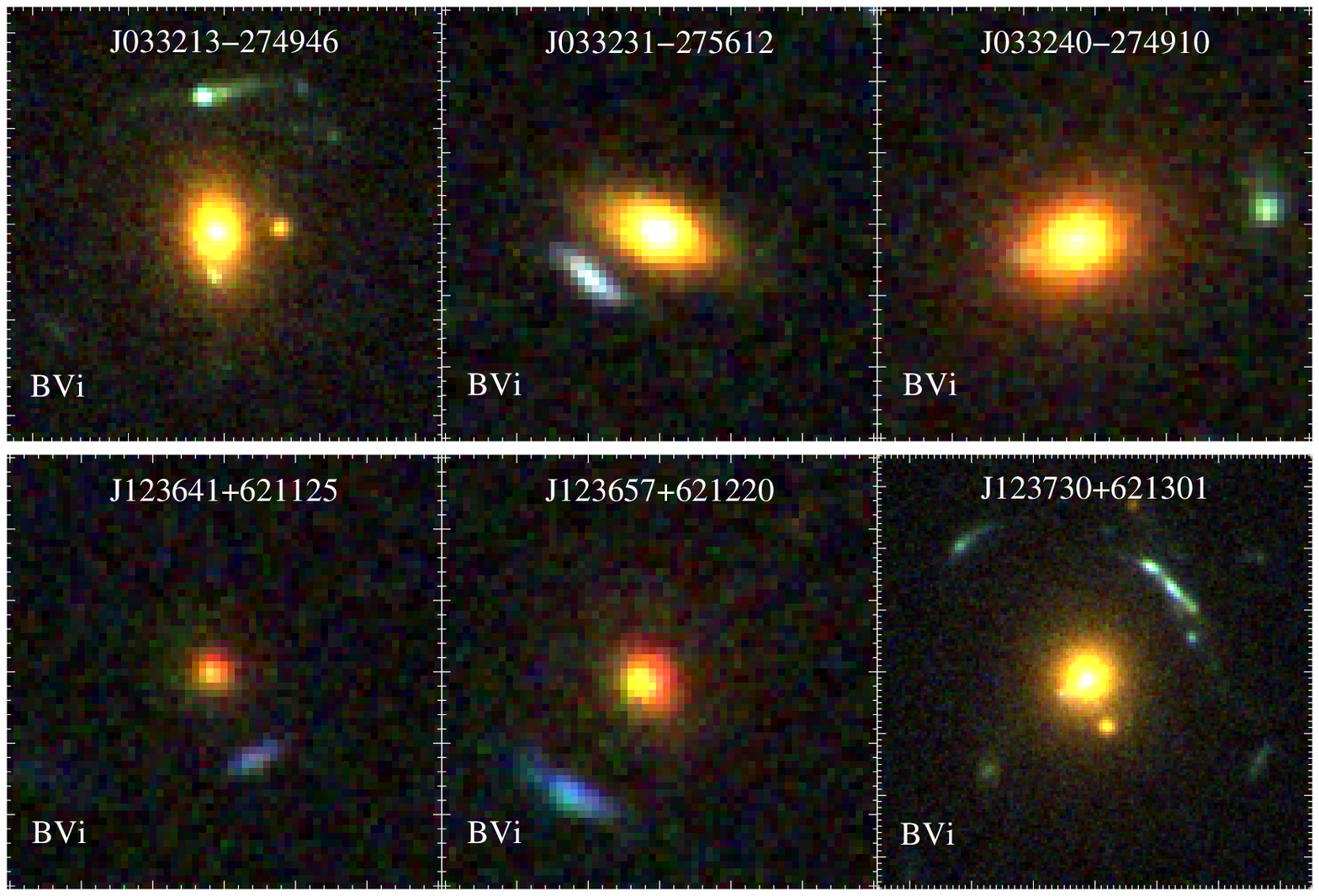}
  \figcaption{Plots showing the $B_{435}V_{606}i_{775}$ 
  images for our top six lens candidates.  All plots are 3\arcsec\ on a side
  with the exception of the plots for J033213$-$274946, which is 4.5\arcsec\ on
  a side, and J123730+621301, which is 7\arcsec\ on a side..
  \label{fig_cand1}}
  \end{figure*}
\fi

\section{Discussion}

The systematic search produced a list of 48
% from goodcand* lists
candidates of varying quality.  The biggest difficulty in selecting
candidates was to differentiate rings or arcs of star formation from
lensing morphologies.  Most of the candidates have morphologies
consisting of arc- or point-like blue features near red galaxies, with
no clear counterimages; these were accepted as candidates because the
faint counterimages might be below the detection threshold.  All of
the authors voted on which of the 48 candidates were most likely to
be lenses.
%Our six top candidates are presented
%in Table~\ref{tab_goodcand} and Figure~\ref{fig_cand1}.
We list our six top candidates
in Table~\ref{tab_goodcand} and show them in Figure~\ref{fig_cand1}.
Further followup is necessary to establish more definitively which of
the systems are real gravitational lenses.  Therefore, we present the
full list of candidates and their coordinates in
Table~\ref{tab_allcand} so that interested readers can make their own
selection of most promising lens candidates and target selected
systems for additional observations\footnote{See also 
\url{http://www.stsci.edu/science/goods/lensing/Lenscands/lenscands.html} 
for images of the full list of candidates.}.
We note that the northern GOODS field includes the HDF-N area, which
was searched extensively for gravitational lenses.  The two best lens
candidates found in those searches, J123652+621227 and J123656+621221
\citep{hogghdf,zmdhdf}, are detected in the GOODS data.  The former is
fainter than our magnitude cutoff at $z_{850,iso}=23.7$; the latter is
in the target sample and was independently selected as lens candidate
GDS J123657$+$621220 (Figure~\ref{fig_cand1}).

In this letter, we will provide only first-order estimates of the
properties of our six best candidates.  Systems for which later
observations clearly establish the lensing nature will be treated more
fully in forthcoming papers.  
A nearly model-independent quantity that may be derived from lens
system data is the mass of the lensing galaxy within a cylinder of
radius $\theta_E$, the Einstein ring radius.  In cases with clear
multiple images, a good estimate of $\theta_E$ is simply half the
image separation.
%(this is an exact relation for a SIS).  
However, in most of the lens candidates presented in this letter, an
arc is seen, but no counterimage is detected.  Although this presents
a difficulty, we note that in cases where an arc is highly magnified
compared to the counterimage, the arc lies very close to the critical
curve of the lens.  Thus, in these cases the distance from the galaxy
to the arc provides a reasonable estimator of $\theta_E$.  
We have used the $V$-band images to estimate $\theta_E$ for the
candidates because both the blue and red objects are detected at high
significance in these images.  In Table~\ref{tab_goodcand} we provide
estimated Einstein ring radii ($\theta_{E,est}$) as well as the target
galaxy magnitudes in apertures with radii $\theta = \theta_{E,est}$.
More extensive descriptions of the properties of the lens galaxies will 
become possible with spectroscopic data and more detailed modeling.

\section{Summary and Future Work
 \label{summary}
}

In this paper we have presented the results of a systematic search for
strong gravitational lens candidates in the GOODS ACS data.  The
search targeted massive early-type galaxies in the redshift range
$0.3<z<1.3$, i.e., the sources most likely to produce strong lenses.
We have presented a list of 48 lens candidates as well as images of
our six most promising candidates.  The search was conducted on images that are
not yet at the full depth expected from the GOODS survey; however, we
do not expect the $\sim$30\% increase in sensitivity provided by the
incorporation of the last two epochs of the GOODS observations to
yield a significant increase in the number of lens candidates.  The
best candidates are being targeted for spectroscopic observations, in
order to confirm or reject the lensing hypothesis.  Based on these
data, we will be able to measure accurate masses, mass-to-light
ratios, and other parameters of the confirmed lens systems.

%%%%%%%%%%%%%%%%%%%%%%%%%%%%%%%%%%%%%%%%%%%%%%%%%%%%%%%%%%%%%%%%%%%%%%%

\acknowledgments 

The GOODS project is very much a team effort, and we would like to
thank all the members of the collaboration, and especially Anton Koekemoer
for his help, patience, and goodwill.  We are grateful to Lori Lubin
and Chris Conselice for useful discussions and to Jeff Valenti for his
%willingness to share his 
IDL expertise.  We thank the referee for
comments that improved the paper.
We are indebted to the shuttle astronauts who, as we have
been so painfully reminded, risk their lives so that we can study the
Universe.  Support for this work was provided by NASA through grant
GO09583.01-96A from the Space Telescope Science Institute, which is
operated by the Association of Universities for Research in Astronomy,
under NASA contract NAS5-26555. Support for this work, part of the
{\em Space Infrared Telescope Facility (SIRTF)} Legacy Science
Program, was provided by NASA through contract number 1224666 issued
by the Jet Propulsion Laboratory, California Institute of Technology
under NASA contract 1407.

%%%%%%%%%%%%%%%%%%%%%%%%%%%%%%%%%%%%%%%%%%%%%%%%%%%%%%%%%%%%%%%%%%%%%%%

\ifsubmode
  \newpage
\fi

\ifsubmode
  \clearpage
\fi

\begin{deluxetable}{lrrrrrrl}
\tablewidth{0pt}
%\tabletypesize{\scriptsize}
\tablecaption{Lens Candidates\label{tab_allcand}}
\tablehead{
\colhead{ID}
 & \colhead{RA(2000)}
 & \colhead{Dec(2000)}
 & \colhead{$B_{435iso}$}
 & \colhead{$V_{606,iso}$}
 & \colhead{$i_{775,iso}$}
 & \colhead{$z_{850,iso}$}
 & \colhead{$z_{phot}$\tablenotemark{a}}
}
\startdata
GDS J033206$-$274729 & 03 32 06.431 & $-$27 47 28.76 & 23.83 & 22.94 & 21.78 & 21.09 & 0.96 \\    %s-v1.7-16475
GDS J033211$-$274650 & 03 32 11.403 & $-$27 46 49.98 & 24.42 & 22.77 & 21.70 & 21.34 & 0.46 \\    %s-v1.7-16769
GDS J033213$-$274946 & 03 32 13.006 & $-$27 49 46.08 & 24.02 & 21.98 & 20.65 & 20.19 & 0.55 \\    %s-v1.7-12124
GDS J033215$-$274157 & 03 32 14.825 & $-$27 41 57.17 & 25.38 & 23.60 & 22.37 & 22.00 & 0.55 \\    %s-v1.7-24097
GDS J033216$-$274714 & 03 32 15.805 & $-$27 47 13.61 & 24.44 & 22.44 & 20.98 & 20.51 & 0.62 \\    %s-v1.7-15592
\enddata
\tablecomments{Units of right ascension are hours, minutes, and seconds.  
Units of declination are degrees, arcminutes, and arcseconds.
[The complete version of this table is in the electronic edition of the Journal.
The printed edition contains only a sample.]}
\tablenotetext{a}{Photometric redshifts determined as described in
\citet{goodsphotoz}.  
Photometric redshifts for the northern 
GOODS sources 
%in the northern GOODS field 
are being determined.}
\end{deluxetable}

\begin{deluxetable}{lrrlrrrrrl}
\tablewidth{0pt}
\tabletypesize{\scriptsize}
%\footnotesize
\tablecaption{Parameters of Candidate Lens Galaxies\label{tab_goodcand}}
\tablehead{
\colhead{ID}
 & \colhead{RA(2000)}
 & \colhead{Dec(2000)}
 & \colhead{$z_{phot}$\tablenotemark{a}}
 & \colhead{$\theta_{E,est}$\tablenotemark{b}}
 & \colhead{$B_{435,E}$\tablenotemark{c}}
 & \colhead{$V_{606,E}$\tablenotemark{c}}
 & \colhead{$i_{775,E}$\tablenotemark{c}}
 & \colhead{$z_{850,E}$\tablenotemark{c}}
 & \colhead{Notes\tablenotemark{d}}
}
\startdata
GDS J033213$-$274946  & 03 32 13.006 & $-$27 49 46.08 & 0.55    & 0.96 & 24.82 & 22.35 & 20.93 & 20.48 & Blob + possible CI \\ %s-v1.7-12124  
GDS J033231$-$275612  & 03 32 31.067 & $-$27 56 12.34 & 0.57    & 0.62 & 25.00 & 23.10 & 21.82 & 21.42 & Arc \\ %s-v1.7-00785  
%GDS J033236$-$274812  & 03 32 35.961 & $-$27 48 11.87 & 0.44    & 0.40 & 24.37 & 22.96 & 22.01 & 21.79 & \\ %s-v1.7-11813  %
GDS J033240$-$274910  & 03 32 39.598 & $-$27 49 09.57 & 0.75    & 1.31 & 25.52 & 23.18 & 21.69 & 20.80 & Faint CI \\ %s-v1.7-10001  
%GDS J033242$-$275157  & 03 32 41.637 & $-$27 51 57.44 & 0.75    & 1.23 & 24.70 & 22.65 & 21.13 & 20.53 & \\ %s-v1.7-05425  %
%GDS J123610$+$620845  & 12 36 09.581 & $+$62 08 45.12 & \nodata & 0.39 & 24.90 & 23.56 & 22.25 & 21.68 & \\ %n-v1.7-00862  %
%GDS J123625$+$621301  & 12 36 25.058 & $+$62 13 01.01 & \nodata & 1.07 & 24.10 & 22.15 & 20.99 & 20.60 & \\ %n-v1.7-08761  %
GDS J123641$+$621125  & 12 36 41.139 & $+$62 11 25.32 & \nodata & 0.73 & 26.82 & 24.95 & 23.57 & 22.73 & Faint arc \\ %n-v1.7-09793  
GDS J123657$+$621220  & 12 36 56.642 & $+$62 12 20.45 & \nodata & 1.00 & 25.77 & 24.25 & 22.71 & 21.97 & Arc \\ %n-v1.7-13652  
GDS J123730$+$621301  & 12 37 29.900 & $+$62 13 01.15 & \nodata & 1.98 & 23.10 & 21.29 & 20.22 & 19.85 & Multiple arcs (+CI?) \\ %n-v1.7-19788
\enddata
\tablecomments{Units of right ascension are hours, minutes, and seconds.  
Units of declination are degrees, arcminutes, and arcseconds.}
\tablenotetext{a}{Photometric redshifts determined as described in
\citet{goodsphotoz}.  
Photometric redshifts for the northern GOODS sources 
are being determined.}
\tablenotetext{b}{Estimated Einstein ring radius, in arcseconds.}
\tablenotetext{c}{Magnitude within an aperture of radius $\theta_{E,est}$.}
\tablenotetext{d}{CI = counter-image.}
\end{deluxetable}

\ifsubmode
  \clearpage
  
  \begin{figure}
  \plotone{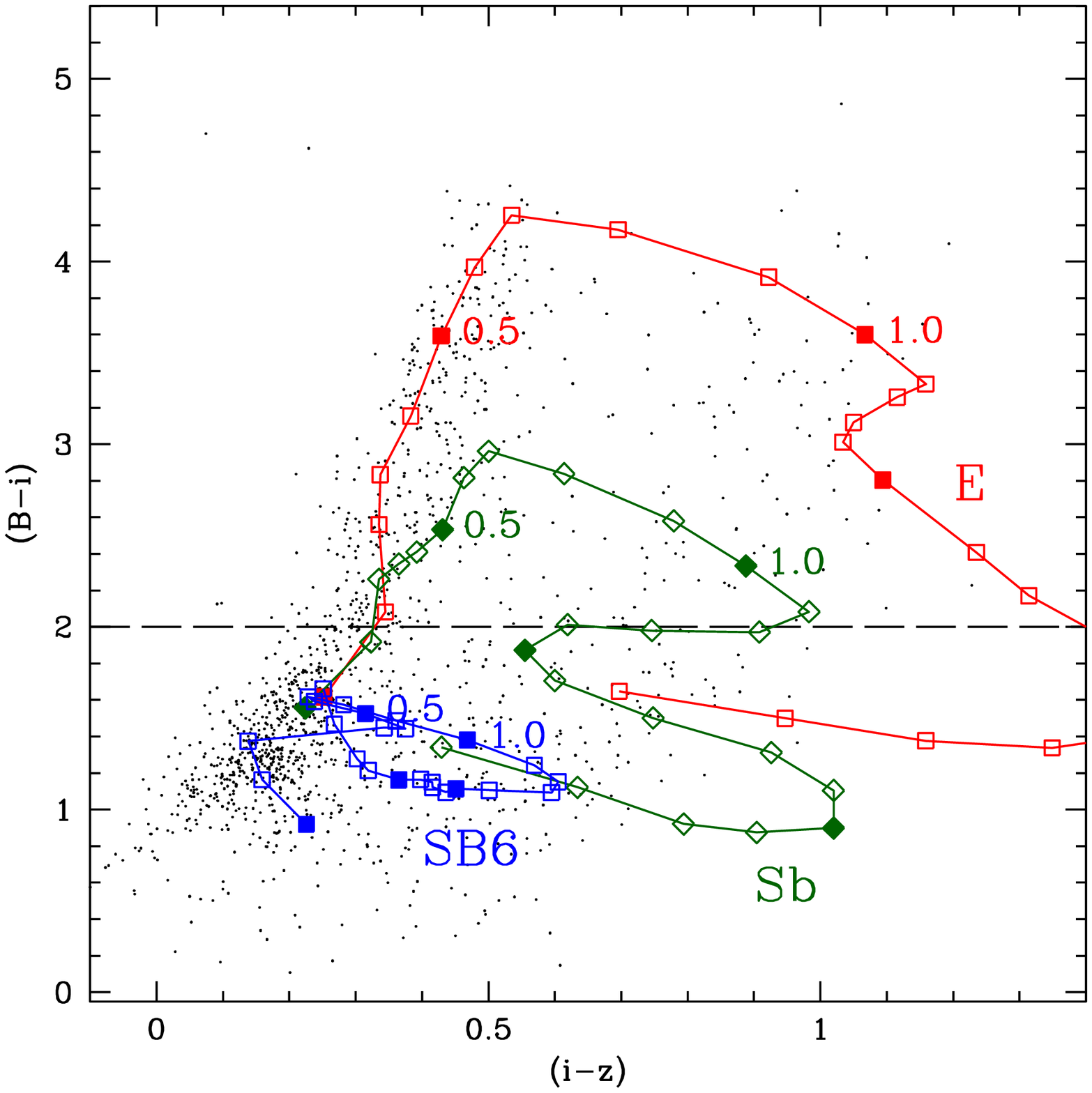}
  \caption{
   Color-color diagram for galaxies in the GOODS southern field
   $z_{850}$-selected catalog, limited to $z_{850,iso}<23.0$.  The target
   galaxies for the lens search lie above the dashed line.  The open
   symbols connected by lines represent the expected change with redshift
   of colors for non-evolving elliptical (E), Sb, and starburst (SB6)
   galaxies.  The points are separated by $\Delta z = 0.1$, starting at
   $z = 0$ and with solid points every $\Delta z = 0.5$.  On each track,
   the points corresponding to $z=0.5$ and $z=1.0$ are marked.
   % The plot for the northern GOODS field is qualitatively similar.
  \label{fig_ellselect}}
  \end{figure}

  \begin{figure}
  \plotone{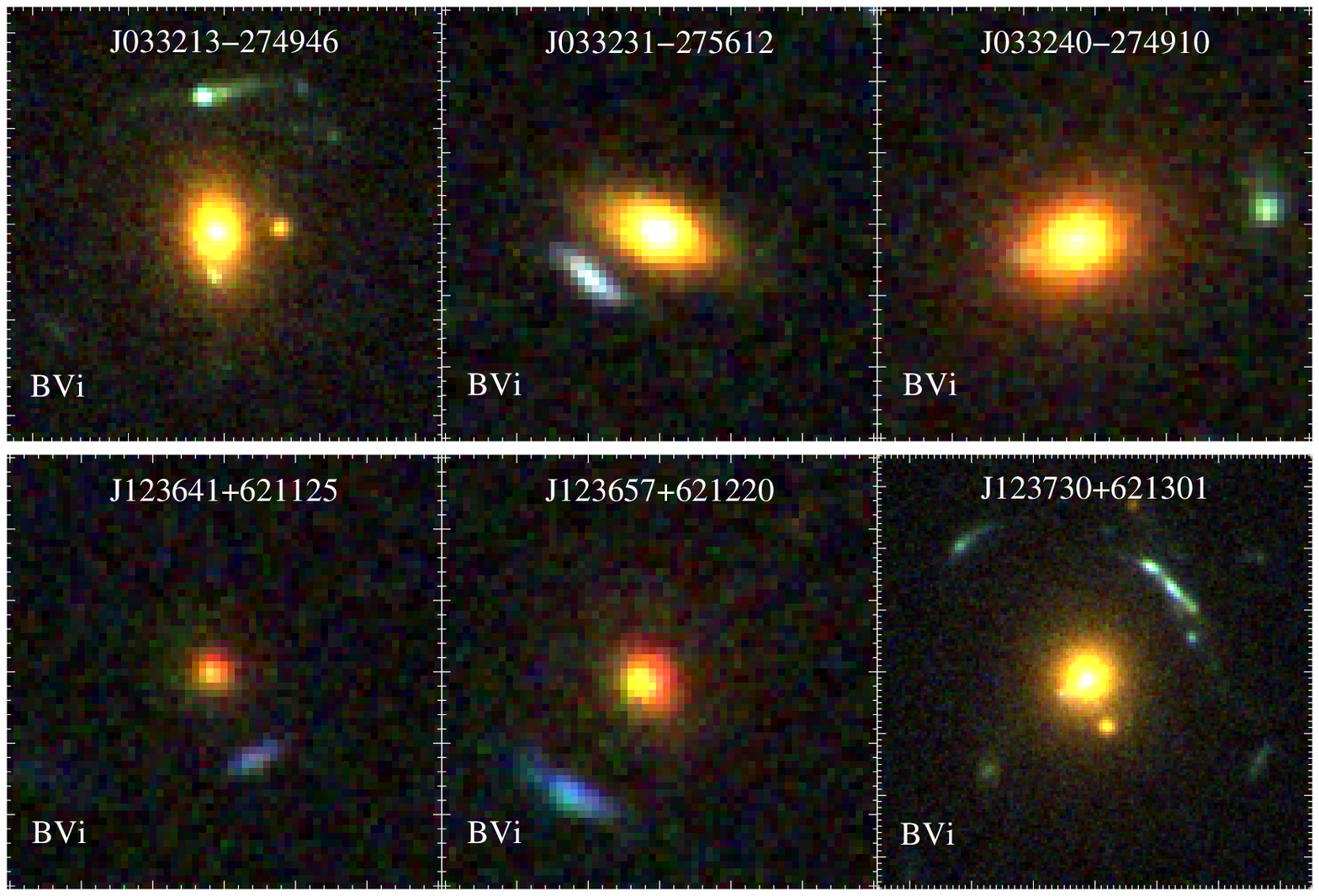}
  \figcaption{Plots showing the $B_{435}V_{606}i_{775}$ 
  images for our top six lens candidates.  All plots are 3\arcsec\ on a side
  with the exception of the plots for J033213$-$274946, which is 4.5\arcsec\ on
  a side, and J123730+621301, which is 7\arcsec\ on a side..
  \label{fig_cand1}}
  \end{figure}

\fi

\end{document}